\documentclass[12pt]{article}
\usepackage{amsmath,amsthm,amsfonts,amssymb}
\usepackage{graphicx}
\usepackage{enumerate}
\usepackage{booktabs}
\usepackage{natbib}
\usepackage{url} 
\usepackage{array}
\usepackage{amscd}
\usepackage{color}
\usepackage{mathrsfs}
\usepackage{latexsym}
\usepackage{bm,bbm}
\usepackage{extarrows}
\usepackage[colorlinks=true, linkcolor=blue, citecolor=blue]{hyperref}
\usepackage{algorithm}
\usepackage{algpseudocode}

\usepackage{colortbl,xcolor,multirow}
\newcolumntype{f}{>{\columncolor{lightgray}}l}
\newcolumntype{h}{>{\columncolor{lightgray}}r}

\begin{document}

\title{\bf A Note on Threshold Principle, Artificial Neural Network Connections  and Econometric Legacy}

  \author{Howell Tong\\
    Paula and Gregory Chow Institute for Studies in Economics, \\
    Xiamen University, China \\
    Department of Statistics and Data Science
    Tsinghua University, China and\\
Department of Statistics, London School of Economics, United Kingdom}

\maketitle

\begin{abstract} 
\noindent This is mostly a non-technical note, reflecting the author's philosophy and personal views on time series analysis, with references limited to only a few representatives for brevity. The Threshold Principle, formally announced in Tong (1990), modernised time series analysis by introducing a collection of  sub-systems  to model complex nonlinear dynamics. We explore the conceptual architecture of the Threshold Principle drawing parallels and contrasts with artificial neural network in machine learning. We trace the influential adoption of Threshold Autoregression in econometrics.
We examine a smooth extension, namely the smooth threshold  autoregression introduced by Chan and Tong (1986)
 that was later popularized in the Econometric literature.  We sound cautions to help econometric users to avoid misuse of this model.  
Furthermore, we examine how we can  
systematically apply the Threshold Principle to conditional variance to enable meaningful volatility classification. Finally, we mention some of the modern applications of the Threshold Principle to non-real-valued domains, underscoring its enduring half-century methodological significance as embodied in the threshold autoregression. 
\end{abstract} 

\vspace{1.5em} 

,
\noindent
{\it Keywords:}
Threshold principle, Threshold autoregression, Smooth threshold autoregression, Dynamical system, Nonlinear time series, Artificial Neural network, Machine learning, Conditional computation, Volatility, Heteroscedasticity, ARCH model, Econometrics, Economics, Finance, Non-real-valued data.

\section{The Threshold Principle} 
More than 35 years ago, the \textbf{Threshold Principle (TP)} was announced (Tong, 1990) \textit{that allows the analysis of a complex stochastic system by decomposing it into simpler subsystems.} 
When applied to nonlinear time series analysis, the paradigm produces the threshold autoregression (TAR), and asserts that \textit{local linearity across regimes separated by thresholds is a powerful way to model global dynamic complexity.} In practice, by decomposing the state space of a complex, global nonlinear dynamical system into regimes (via thresholds), each of which is governed by linear dynamics, the TAR, and by implication TP, enables time series models to capture rich nonlinear phenomena such as asymmetry, limit cycles, chaos, jump phenomena and many others (Tong, 1990). 

We note that, for time series, TAR is not the only way to implement the TP but we focus on TAR in this note as it was historically the first. It should be noted that the TP also applies to continuous-time systems, although this note focuses on discrete time. For further details on continuous-time TAR, see, e.g., Tong (1990) and Su and Chan (2015) .    Further,  the TP also applies to static systems in which time is absent, for example, two-phase regression (Quant, 1958), Tukey's regressogram (1961), threshold panel regression (Hansen,1999) and others. The first two precedated the TAR. To put things into proper  perspective, recall that though Yule's autoregression was predated by the age-old regression, Yule's recognition of the role of time in 1927 was most significant, almost as significant as Einstein's $t$ in his 4-space $(x,y,z,t)$ in 1905. Likewise, without time, threshold models cannot capture the multitude of nonlinear dynamical phenomena mentioned previously. In short, without TIME, the world is still. To some, including the author, it is not as exciting. 

\section{Philosophical Similarities and Technical Differences: TP/TAR vs Artificial Neural Network} 

As the earliest product of the TP,
we denote typically a two-regime TAR  of order $p$ for time series $\{Y_t: t=0,1,2,...\}$  as

\begin{equation} 
Y_t = \left( \phi_{1,0} + \sum_{i=1}^p \phi_{1,i} Y_{t-i} \right) I(Z_t \le r) + \left( \phi_{2,0} + \sum_{i=1}^p \phi_{2,i} Y_{t-i} \right) I(Z_t > r) + \epsilon_t,
\end{equation} 
where $Z_t$ is the threshold variable (e.g., $Y_{t-d}$), $r$ is the threshold parameter, $I(\cdot)$ is the indicator function taking value 1 if the condition holds and 0 otherwise, and $\epsilon_t \sim \text{i.i.d.}(0, \sigma^2)$.

Below is  Artificial Neural Network (ANN) in its basic form:
For a single neuron $j$ receiving inputs $x_i$ with weights $w_{ij}$ and bias $b_j$, and outputting $y_j$:

\begin{equation}
    z_j = \left( \sum_{i} w_{ij} x_i \right) + b_j,
\end{equation}

\begin{equation}
    y_j = f(z_j),
\end{equation}
where the activation function $f$ (historically the logistic sigmoid function) is defined as:

\begin{equation}
    f(z) = \frac{1}{1 + e^{-z}}.
\end{equation}
See
Cheng and Titterington (1994) for a review of ANN from a statistical perspective.
\\

The TAR shares profound conceptual roots with ANN, and yet the two frameworks diverge in execution and analytical intent. Both paradigms reject global linearity in favour of localized linear approximations. However, while ANN prioritizes flexible function approximation in high-dimensional spaces, TP prioritizes structural interpretability and statistical inference, although, as shown by Petruccelli (1992), TAR can provide an almost sure approximation of general classes of time series models.

\begin{table}[h!] \centering \small \caption{Comparison between Threshold Principle and Artificial Neural Network} 
\vspace{0.5em} 
\begin{tabular}{p{3.2cm} p{6.0cm} p{6.0cm}} 
\toprule 
\textbf{Dimension} & \textbf{Threshold Principle (TP / TAR)} & \textbf{Artificial Neural Network (ANN)} \\ \midrule \textbf{Philosophical Goal} & Piecewise linear approximation of complex global dynamics. & Superposition of localized linear transformations via activation gates. \\ 
\addlinespace \textbf{Space} & \textbf{Explicit:} Partitioned by  threshold variables and boundaries. & \textbf{Implicit \& Distributed:} Hidden layers form possibly soft hyperplanes in high dimensions. \\ 
\addlinespace \textbf{Interpretability} & \textbf{High Structural Rigor:} Clear regime definitions (e.g., expansion vs. recession). & \textbf{Black-Box:} Complex distributed representations with low direct interpretability. \\ \addlinespace \textbf{Estimation} & Conditional Least Squares, Grid Search. & Backpropagation, Stochastic Gradient Descent (SGD). \\ 
\addlinespace 
\textbf{Data Requirements} & Small-to-moderate sample sizes in low dimension/medium dimension. & Large-scale data required to fit multi-layer parameters in high dimension. \\ 
\bottomrule 
\end{tabular} 
\end{table}

The difference is particularly striking. 
The central idea of TAR is both simple and profound: a time series may obey different autoregressive laws in different regions (called regimes)  of its state space, with the transition between regimes determined by a threshold.
 An ANN seeks to learn a flexible nonlinear function from the data, whereas TAR explicitly identifies a structural feature of the process—a threshold at which the underlying dynamics change. Thus, while an ANN may reproduce a nonlinear behaviour/function very effectively, TAR offers a statistical interpretation of \textit{\textbf{why}} that behaviour changes. (In fact, we can say that the ANN represents a black-box approach while TAR a grey-box approach.) 
The subsequent development of the smooth threshold autoregression (STAR), notably through the work of Chan and Tong (1986), replaced the abrupt TAR transition by a gradual one. In this sense, we can view the progression conceptually as

\vspace{0.5cm}
\noindent
\textbf{AR → TAR → STAR → increasingly flexible nonlinear models such as ANN.}

\vspace{0.5cm}

\noindent This should not, however, be interpreted as a direct historical genealogy of ANN. Rather, it represents increasing flexibility in the representation of nonlinear dynamics.

The enduring significance of TAR, as a direct product of TP, is therefore not simply that TAR provided a better prediction technique than the linear approach. It introduced a new way of thinking about time series: \textit{nonlinearity can arise because different dynamical mechanisms operate in different regimes of the process}. The threshold becomes a {\it scientifically meaningful feature} of the system rather than merely a device for improving goodness of fit and prediction.

In this respect, TAR anticipated an important theme of modern machine learning, namely conditional computation (CC). CC   addresses  something that highly flexible black-box models often sacrifice, namely \textit{interpreability and a clear statistical structure}. See, e.g., Zhou {\it et al.} (2022).

\section{TAR for Econometrics and Finance}

While originating in statistics, interestingly (and surprisingly to the author) the TAR found its most significant empirical applications in 
econometrics and finance. This is undoubtedly due to the fact that TAR is ideally suited to model  {\it regime-switching} that is significant in economics and finance. For, economic and financial activities, such as transaction costs, policy target bands, and business cycle asymmetries, naturally exhibit regime-switching behaviour. When appropriately fitted, the TAR is found to be ideally suited to produce interpretable and realistic results. 

In his comprehensive review, 
Hansen (2011) systematically documented how econometrics came to account for the largest usage of TAR. He surveyed over 75 core publications spanning theoretical developments (such as testing under non-identified nuisance parameters and estimated thresholds' asymptotic distributions) and extensive empirical applications across macroeconomics, monetary economics, and international finance. In the same issue, Chen et al. (2011) gave a comprehensive review of TAR in finance. These two papers have solidified the TAR framework as a dominant nonlinear workhorse in applied econometric and finance modelling. 

\section{Smooth Threshold Autoregression: Specifications and Technical Caveats} 

Transitions between regimes can be smooth in some cases. Chan and Tong (1986) realised this forty years ago and  introduced the \textit{Smooth Threshold Autoregression} (STAR), by replacing  indicator functions (representing hard thresholds)  with continuous functions (representing  soft thresholds)\footnote{Perhaps a better name is {\it soft threshold autoregression,} retaining the same acronym.} $G(Z_t; \gamma, r) \in [0, 1]$: 
\begin{equation} 
Y_t = \left( \phi_{1,0} + \sum_{i=1}^p \phi_{1,i} Y_{t-i} \right) \left[ 1 - G(Z_t; \gamma, r) \right] + \left( \phi_{2,0} + \sum_{i=1}^p \phi_{2,i} Y_{t-i} \right) G(Z_t; \gamma, r) + \epsilon_t.
\end{equation} 
Several years later, Teräsvirta (1994) and others popularized STAR in empirical econometrics, focusing primarily on two functional forms for $G(Z_t; \gamma, r)$: 
\subsection*{Logistic STAR (LSTAR)} :

\begin{equation} G(Z_t; \gamma, r) = \left( 1 + \exp\left\{ -\gamma (Z_t - r) \right\} \right)^{-1}, \quad \gamma > 0 
\end{equation} 

\subsection*{Exponential STAR (ESTAR)} :

\begin{equation} 
G(Z_t; \gamma, r) = 1 - \exp\left\{ -\gamma (Z_t - r)^2 \right\}, \quad \gamma > 0 
\end{equation} 
Note that LSTAR was inspired by TAR and ESTAR by EXPAR; see Tong (1990).

\subsection*{Critical Cautions in STAR Estimation}
In view of the fact that LSTAR and ESTAR have been fitted in the literature, often uncritically,  to many economic time series, it seems appropriate to repeat a couple of cautionary remarks implicit in Chan and Tong (1986). 

\subsubsection*{Caution (i):  Data Hunger of the Smoothing Parameter} 
It is intuitively obvious that a large sample is needed to get a reasonable estimate of the
’twisting part’ of a continuous function. Now, the smoothing parameter $\gamma$ governs the
twisting part related to the speed of transition between regimes. Estimating $\gamma$ reliably
demands dense data concentrations in the immediate neighbourhood of the threshold $r$.
For samples of moderate size, the likelihood surface with respect to $\gamma$ tends to be quite
flat, leading to imprecise standard error estimates—a critical issue often scarcely addressed by
applied econometric users. Especially for data that is bimodally distributed, we should be particularly vigilant of pitfalls in fitting an STAR if the estimated $r$ is in the vicinity of the anti-mode. The Canadian lynx data is a case in point; see Tong (1990, pages 362 and 363).

\subsubsection*{Caution (ii): Stochastic vs. Deterministic Limit Reductions} It is commonly asserted that a STAR model simply "reduces" to a TAR model as the smoothing parameter $\gamma \to \infty$. While true in a deterministic setting (with the noise term set to a constant, e.g. zero),  in a stochastic setting (where process components are random variables  governed by probability measures), this reduction requires subtle probabilistic conditions. As stated by Chan and Tong (1986), establishing convergence from STAR to TAR necessitates proving \textit{uniform boundedness and equicontinuity} of the underlying stochastic operators to guarantee weak convergence---a nuance frequently overlooked in econometric and statistical literature. 

For an in-depth study,  we refer to Gao {\it et al.} (2018), which includes tests of TAR vs STAR. 

\section{Volatility Modelling with Discrete Classification by T-CHARM} 

Historically speaking, Moran (1953) was possibly the first statistician who paid attention to  heteroscedasticity in time series. Now, besides modelling the trend (the conditional mean) dynamics, the TP provides a simple way to model heteroscedasticity (the conditional variance). In fact, heteroscedasticity was already built into TAR (Tong, 1980) before Engle (1982), who introduced the ARCH model. However, a full methodological development of TAR for conditional variance was rather slow in the making and had to wait till  Chan  {\it et al.} (2014), who introduced the \textit{T-CHARM} (Threshold Conditionally Heteroscedastic Model) framework. By viewing volatility as conditional variance that follows a state-dependent process delineated by threshold variables, T-CHARM  allows the volatility of time series $\{X_t\}$ to vary from one regime to another in a {\it piecewise constant} manner. Briefly, let the real-line be the union of $m$ disjoint intervals, $R_i, \: i=1,2,...,m$ and let $\sigma_i,  \: i=1,2,...,m$ be constants. We write the T-CHARM as
\begin{equation} 
X_t  =  \sigma_t\eta_t = \sigma(X_{t-1})\eta_t, 
\end{equation} 
where $\sigma_t = \sigma_j$ if $X_{t-1}$ lies in $R_j$, and $\{\eta_t\}$ is a sequence of independent and identically distributed random variables, each with zero mean and unit variance.

T-CHARM provides a clear {\it discrete} classification into, for example with $m = 3$, low, medium, and high  volatilities. We argue that, in practice, a discrete classification is often all that is relevant in finance and possibly other areas, rather than a continuous one.  More crucially, as proven by Chan {\it et al.} (2014), the T-CHARM  possesses strict stationarity and ergodicity essentially {\it without} any restriction on the parameters, a feature not shared by the ARCH model or its descendents. Further, to cope with different applications, it has been found necessary to extend the basic ARCH mixing function to ones of increasing complexity, associated with the acronyms GARCH, EGARCH, IGARCH, NGARCH, GARCH-M, QGARCH,
GJR-GARCH, TGARCH, fGARCH, and others. These extensions carry  with them additional parameter constraints.  We would argue quite strongly that there is no {\it a priori} 
reason why the mixing function $\sigma(X_{t-1})$ should be wedded to a continuous form.

We hope that  the simplicity and numerous advantages  detailed in Chan {\it et al.} (2014) might, in future, charm  econometricians and financial engineers into T-CHARM. 

\section{Generalization Beyond Real-Valued Data} 
It is well known that data can present themselves in many different forms in a modern society like ours. Fortunately, although initially applied to real-valued time series, the TP/TAR applies far beyond that. An incomplete list is given below as an indication:  

\begin{itemize}
\item \textbf{Count Data:} Integer-valued TAR processes for epidemic tracking, trade frequencies, and transaction counts. See, e.g., Ma {\it et al.} (2025) and the references therein.
\item \textbf{Panel/Matrix Data:} Panel TAR models capturing cross-sectional interactions across economic networks. See, e.g., Hansen (1999) and Liu and Chen (2022).
\item \textbf{High dimensional data:} High dimensional factor models that switch dynamics across regimes. See, e.g.,   Liu \& Chen (2020).
\item \textbf{Functional Data:} Threshold processes operating on continuous curves, such as yield curves or intraday high-frequency financial paths. See, e.g., Li {\it et al.} (2024).
\item \textbf{Spatial Data:} threshold models for spatial econometrics data. See, e.g., Li and Lin (2024). 
\end{itemize} 

\section{Conclusion: The Half-Century Horizon} The TAR model has served econometrics effectively for over half a century. As documented by Hansen (2011) and Chen {\it et al.}, its  footprint in econometrics and finance is vast. By bridging deterministic dynamical systems theory and stochastic modelling, TP transformed empirical time series analysis.  As high-dimensional, functional, and non-Euclidean data become prevalent, the TP remains poised to guide methodological innovation for decades to come. 

\section*{Acknowlegements} I thank Professors Yongmiao Hong and Shouyang Wang for their encouragements, and Professors  Kung-Sik Chan and Bing Cheng for comments and suggestions. I alone remain responsible for all the errors or misrepresentations. 

\section*{References}
\begin{description}
    \item 
\item Chan, K. S., \& Tong, H. (1986). On estimating thresholds in autoregressive models. \textit{J. Time Series Analysis}, 7(3) 179-190.
\item Chen, C. W. S., So, M. K. P., \& Liu, F. (2011). A review of threshold time series models in finance,  \textit{Statistics and Its Interface}, 4(2), 167-181.
\item Cheng, B, \& Titterington, D. M. (1994) Neural networks: A review from a statistical perspective. \textit{Statistical Science}, 2-30.

\item Engle, R. F. (1982). Autoregessive conditional heteroscedasticity with estimates of the variance of the United Kingdom inflation. \textit{Econometrica}, 50, 987--1007. 
\item Gao, Z., Ling, S. \& Tong, H. (2018). Tests for TAR models vs STAR models--a separate family of hypotheses approach. \textit{Statistica Sinica}, 28(4), 2857--2883. 
\item Hansen, B. E. (1999). Threshold effects in non-dynamic panels: Estimation,testing, and inference. \textit{J. Econometrics}, 93(2), 345-368.
\item Hansen, B. E. (2011). Threshold autoregressions in economics. \textit{Statistics and Its Interface}, 4(2), 123--127. 
\item Li, Y., Chen, K., Zheng, X., \& Yau, C. Y. (2024). Functional threshold autoregressive model. \textit{Statistica Sinica}, 34(2), 817-836.
\item Li, K. \& Lin, W. (2024). Threshold spatial autoregressive model. \textit{J. of Econometrics}
244(1), 105841. 
\item Liu, X. \& Chen, R, (2020). Threshold factor models for high-dimensional time series. \textit{J. of Econometrics}, 216(1), 51-70.
\item Liu, X. \& Chen, E. Y. (2022). Identification and estimation of threshold matrix-variate factor models. \textit{Scan. J. Statist.}, 49(4), 1383-1417
\item Ma, X., Li, D. \& Tong, H. (2025) Modelling time series of counts with hysteresis. \textit{arXiv:2509.15508} (accepted by Statistica Sinica).
\item Moran, P. A. P. (1953). The statistical analysis of the Canadian Lynx cycle. I. Structure and Prediction. \textit{Australian J. Zoology}, 1(2), 163-173.
\item Petruccelli, J. D. (1992). On the approximation of time series by threshold autoregressive models. \textit{Sankhya: The Indian Journal of Statistics Series B}, 54, 106-113.\item Quant, R. E. (1958). The estimation of the parameters of a linear regression system obeying two separate regimes. \textit{J. Amer. Stat. Ass.}, 52(284), 873-880. 
\item Su, F. \& Chan, K. S. (2015). Quasi-likelihood estimation of a threshold diffusion process. \textit{J. Econometrics}, 189(2), 473-484.
\item Teräsvirta, T. (1994). Specification, estimation, and evaluation of smooth transition autoregressive models. \textit{J. Amer. Stat. Ass.}, 89(425), 208--218. 
\item Tong, H. (1980). A view on non-linear time series model building. \textit{Time Series, Edited by O. D. Anderson, North-Holland Publishing Company}, 41-56.
\item Tong, H. (1990). \textit{Non-linear Time Series: A Dynamical System Approach}. Oxford University Press. 
\item Tukey, J.W. (1961). Curves as parameters, and touch estimation. \textit{Proc. 4th Berkeley Symposium on Mathematical Statistics and Probability}, 1, 681-694.
\item Zhou, Y., Lei, T., Liu, H., Du N., Huang, Y., Zhao, V. Y., Dai, A. M., Chen, Z., Le Q. V., \& Laudon, J. (2022). Mixture-of-experts with expert choice routing. \textit{Google Brain}. 
 
\end{description} 
\end{document}